# Machine-Learning-Based Intelligent Framework for Discovering Refractory High-Entropy Alloys with Improved High-Temperature Yield Strength

Stephen A. Giles[1*], Debasis Sengupta[1*], Scott R. Broderick[2], Krishna Rajan[2]

[1] CFD Research Corporation, 791 McMillian Way, NW, Huntsville 35806
[2] Department of Material Design and Innovation, University at Buffalo, 120 Bonner Hall, NY-14260

* debasis.sengupta@cfd-research.com
* stephen.giles@cfd-research.com

**Abstract**

Refractory high-entropy alloys (RHEAs) are a promising class of alloys that show elevated-temperature yield strengths and have potential to use as high-performance materials in gas turbine engines. However, exploring the vast RHEA compositional space experimentally is challenging, and only a small fraction of this space has been explored to date. The work demonstrates the development of a state-of-the-art machine learning (ML) predictive framework coupled with optimization methods to intelligently explore the vast compositional space and drive the search in a direction that improves high-temperature yield strengths. Our forward yield strength model is shown to have a significantly improved predictive accuracy relative to the state-of-the-art approach, and also provides inherent uncertainty quantification through the use of repeated *k*-fold cross-validation. Upon development of a robust yield strength prediction model, the coupled framework is used to discover new RHEAs with superior high temperature yield strength. We have shown that RHEA compositions can be customized to have maximum yield strength at a specific temperature.

**Introduction**

High-entropy alloys (HEAs) are promising materials, which have garnered a tremendous amount of attention since their discovery in 2004[1–7]. Unlike traditional alloys, HEAs contain at least four elements in near-equal proportions, and the stability of the alloys are postulated to arise from higher configurational entropy[8]. Experimental studies have shown that refractory HEAs (RHEAs) possess superior high-temperature strength compared to superalloys, making them an attractive class of alloys for further exploration[9–15] for potential use in high efficiency gas turbine engines. While HEAs provide tremendous opportunity due to the flexibility of the compositional space, they also pose stiff challenges to the material scientists tasked with exploring a design space with a huge number of possible compositions. Recently, Miracle *et al.* analyzed that using 75 elements that are not toxic, radioactive or noble gas, one can form 219 million 3-6 component base alloys[16]. If elemental composition is varied for each base alloy, the number of new HEAs becomes more than 592 billion[16]. To date, only a tiny fraction of this composition space has been processed and characterized. One of the primary obstacles that impedes the accelerated development of HEA is the lack of generalized understanding of parameters that are responsible for dictating the mechanical and chemical behavior of these complex alloy systems. Although atomic and microstructural information can be obtained with modern high-resolution imaging techniques, these techniques are time-intensive and costly, thereby limiting the extent to which the vast composition space can be explored and characterized.

As a consequence, HEA research has mainly revolved around identifying rules for phase formation and atomic and microstructural parameters potentially affecting mechanical properties and to develop criteria for classification of phases, such as such as yield strength (YS), ductility/plasticity, hardness, and fatigue. Extensive work has been reported in the literature[17–19] to classify phases using parameters, such as the atomic size mismatch ($\delta$), the enthalpy of mixing ($\Delta H_{mix}$), the entropy of mixing, $\Delta S_{mix}$, and dimensionless quantities $\Omega$, $\Phi$, and $\varphi$. Ranges of these parameters have been proposed that can lead to different phases. However, developing guidelines for the improvement of mechanical properties is a relatively less studied area. Earlier reports in this area are primarily concentrated on experimental efforts concerning the modification of a base alloy and finding correlations between mechanical properties with parameters, such as lattice distortion[20–23], grain size[10,24,25], and phases[26–28]. While this approach can be beneficial when narrowly focused on a particular base alloy system, it has limited transferability to other systems. Most recently, Maresca *et al.* have used an edge-dislocation base analytical model and generated ~10 million compositions with increasing theoretical yield strengths[29]. Using their analytical model, a large number of HEAs were discovered that would potentially result in yield strength improvement over the existing HEAs. However, the use of machine learning (ML) models may offer superior accuracy for the prediction of mechanical, phase, and other physicochemical properties. In fact, a number of articles have recently appeared that have applied various ML methods to predict HEA phases[30–32]. However, research on the prediction of mechanical properties of HEAs, in particular in the high-temperature regime, is scarce. In addition, to our knowledge, there is no reported work on developing modeling strategies for intelligently searching the HEA compositional space in a direction that improves mechanical properties.

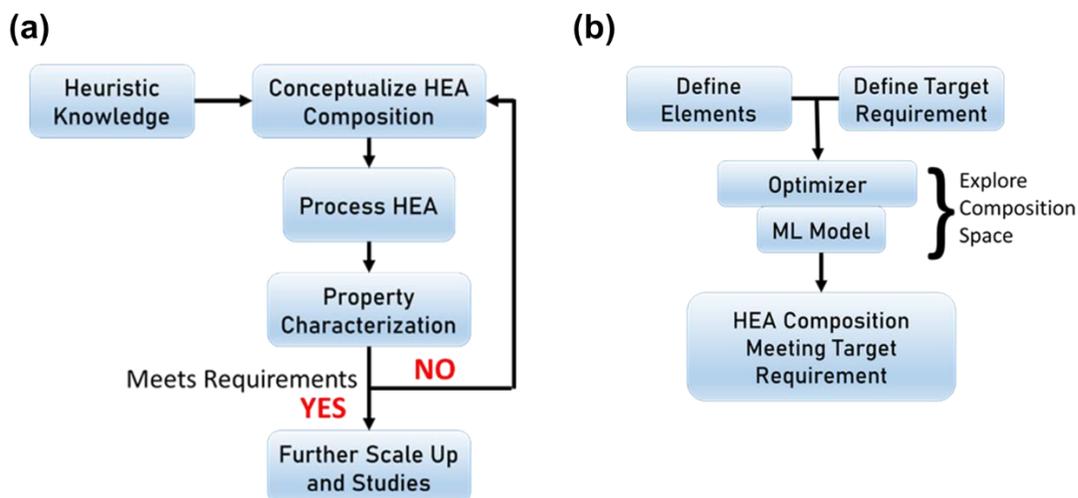

*Figure 1: (a) Traditional HEA processing requires exploring the composition space experimentally to achieve a target property. This makes exploring a large HEA composition space and discovering new HEAs difficult, time-consuming, and expensive; (b) the flowchart of the procedures which explore the composition space intelligently, using optimization coupled with an ML model to achieve a target requirement. The approach narrows down the search space for material scientists potentially accelerating new HEA discovery.*

Currently, HEA development starts with processing a base alloy followed by generating a few HEAs with varying compositions. If the properties of the new compositions did not improve, the search for new alloys continues (Figure 1a). The work presented in this paper aims at substituting this "experiment-only" loop with intelligent machine learning (ML)-based models to screen HEAs and narrow down the search space providing material scientists only a few alloys potentially showing improved properties to process and characterize (Figure 1b). This paper, as an example, focuses on yield strengths of refractory high-entropy alloys (RHEAs) and begins with developing a comprehensive forward ML model *via* identifying critical descriptors selected from a set of a large number of descriptors. The forward model was then coupled with a stochastic genetic algorithm[33] to discover new RHEAs compositions with improved yield strengths. Valuable insights were gained with respect to identifying elements contributing to improving yield strength at room and high temperatures. The physical and thermodynamic descriptors were also analyzed to understand their roles in improved yield strength. Using the ML-based model, we discovered new RHEA compositions with yield strengths customized for specific temperatures. Beginning with known experimental compositions of RHEAs, we have discovered several new compositions with significant improvement in yield strength over the starting compositions.

**Results and Discussion**

The RHEA yield strength data were obtained from the publication by Couzinie *et al.*[34]. The data in the publication contained both compression and tension data. For the present work, only compression data was used as the tension data were insignificant in number. Since the article did not contain detailed processing conditions (e.g., annealing temperature, annealing time, etc.), we extracted the specific processing conditions from the original articles referenced by Couzinie *et al.*[34] The revised dataset contained 284 temperature-dependent yield strength values for unique alloys. Once the data were collected, a large number of composition based descriptors were generated using the *matminer* library[35]. Calculation for several additional composition-based descriptors, scuh as lattice and modulus of distortion, were implemented in *matminer.* Prior to training and validation of the ML models, the descriptor set was reduced by removing descriptors which had either undefined, low-variance, or linearly dependent values. Specific details of our approach are provided in the Methods section. A comprehensive list of the descriptors that were made available for the feature selection and ML training process is provided in the Supplementary Information (SI).

Figure 2a shows the parity plot comparing the measured yield strength values to the predicted values following training and validation using the random forest model[36]. The six descriptors chosen by the sequential feature selection (SFS) method[37,38] are the test temperature, $\Omega$, atomic size mismatch ($\delta$), tantalum modulus distortion ($dG_{Ta}$), fractional composition of molybdenum

($x_{Mo}$), and a base strength, $\sigma_{0,\,min0.5}$, determined from the yield strength of the individual elements in the alloy. The model was found to have a cross-validation regression coefficient, $R^2$ (CV), of 89.5%. All data in the parity plot are colored by the temperature at which the yield strength measurements were made. As expected, the measurements performed above room temperature generally have lower yield strengths than measurements performed at or below room temperature. Our model is shown to provide an excellent quantitative agreement over the entire temperature range. Models were also developed using LASSO regression[39], ridge regression[40], and gradient boosting regression[41]. These models are summarized in Figures S1 – S3 of the SI. LASSO and ridge regression resulted in significantly lower accuracy due to their linear nature, whereas gradient boosting resulted in a $R^2$ (CV) comparable to that of random forest model shown in Figure 2. Our present discussion is therefore based on the results obtained with the random forest model. To understand the effect and importance of each of the six selected descriptors on the predicted yield strength, we have applied the Shapley Additive Explanations (SHAP) technique[42,43] to physically interpret the forward model that was validated in Figure 2a. SHAP analysis is a game-theoretic, local explanation method which computes the quantitative influence of model descriptors on the model output. Although never applied in materials science, SHAP analysis can provide some insights into the model over the entire data space which is otherwise difficult to obtain. As expected, the test temperature is shown via SHAP analysis to be the most important feature, with high temperature corresponding to a lower predicted yield strength and vice versa (Figure 2b). The other important conclusion of the SHAP analysis is that higher values of $\delta$, $\delta G_{Ta}$, and $x_{Mo}$ tend to positively contribute to the yield strength. The outlier red points for $\Omega$ result from smaller values of $|\Delta H_{mix}|$, thereby causing $\Omega$ to become very large (see Equation ((4)). The trend of $\Omega$ versus the SHAP impact on the model by $\Omega$ is shown in Figure S4. The trend reveals that the impact of $\Omega$ undergoes a such that for $\Omega < \sim 15$, $\Omega$ clearly has a positive impact on the yield strength, whereas for $\Omega > \sim 15$, $\Omega$ clearly has a negative impact on the yield strength. Therefore, this model implies the connection between the mechanical property and the parameters derived from atomic structure, and also provides a manner through which to directly link the two.

For the predicted values in Figure 2a that were determined from performing 1,000 *k*-fold cross-validations (see the Method section for details), the standard deviation of the predicted value of each alloy was determined. The distribution of these standard deviations is shown in Figure 2c. The distribution is log-normal due to all values being positive, by definition, and the average standard deviation was found to be 48 MPa. We additionally calculated the error in the predictions relative to the known experimental values. This error distribution is shown in Figure 2d. In this case, the distribution is normal with a mean value near zero, and a characteristic mean *absolute* error (MAE) of 118 MPa. The corresponding percentage error distribution is shown in Figure S5. By contrast, the theoretical model proposed by Varvenne and Curtin[44] was determined to have an MAE of 683 MPa for the same dataset, and consistently underpredicted the experimental yield strength. The corresponding parity plot and error distribution are provided in Figure S6. Therefore, our model represents a significant step forward in the state-of-the-art.

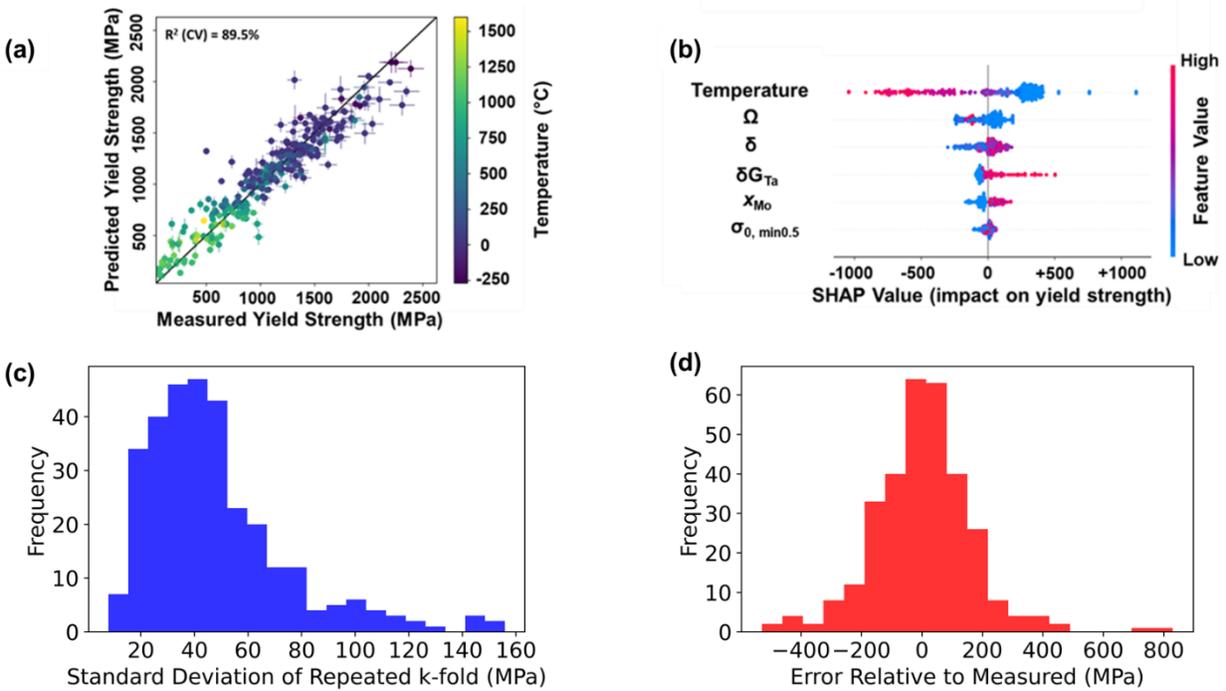

*Figure 2: (a) Predicted vs. experimental yield strength for compression data using the model developed according to the procedures described in the Methods section. Note that we used a repeated k-fold method where 5-fold cross validation was performed 1000 times to compute the mean and standard deviation of each data point; (b) SHAP analysis is performed to identify descriptor importance. Features on the left are ordered according to their importance, with the most important feature (temperature) shown on top. The data points correspond to the individual alloy data points, where each have been colored according to the magnitude (high or low) of the feature in question. Positive SHAP values indicate that the yield strength is increased as a result of the feature value, whereas negative SHAP values indicate the yield strength is decreased due to the feature value. (c) Distribution of prediction standard deviations for each data point when 5-fold cross-validation is performed 1,000 times. (d) Distribution of prediction errors relative to the measured values.*

Figure 3 shows the temperature dependent prediction of four different RHEAs along with their experimental yield strength. Readily apparent from Figure 3 is an excellent agreement between predicted and experimental yield strengths over the entire temperature range. The temperature-dependent yield strength predictions of other RHEAs are provided in the SI. Two of these RHEAs have garnered some attention recently. The AlMo$_{0.5}$NbTa$_{0.5}$TiZr HEA is of interest due to its exceptional high-temperature yield strength (745 MPa at 1,000 °C)[13,45–47]. It has been reported in multiple studies, and is among the state-of-the-art RHEAs. The HfNbTaTiZr alloy is also multiply attested, and is of interest due to its high ductility under tension[48–50]. However, HfNbTaTiZr has a relatively poor yield strength (929 MPa at 25 °C). In the following section, AlMo$_{0.5}$NbTa$_{0.5}$TiZr and HfNbTaTiZr were used as the base alloys for subsequent improvement of the yield strength through compositional optimization. Finally, CrNbTiVZr and MoNbTaTiW were chosen to complete the remaining elements (i.e., Cr, V, and W) that are covered in the Couzinie dataset[34].

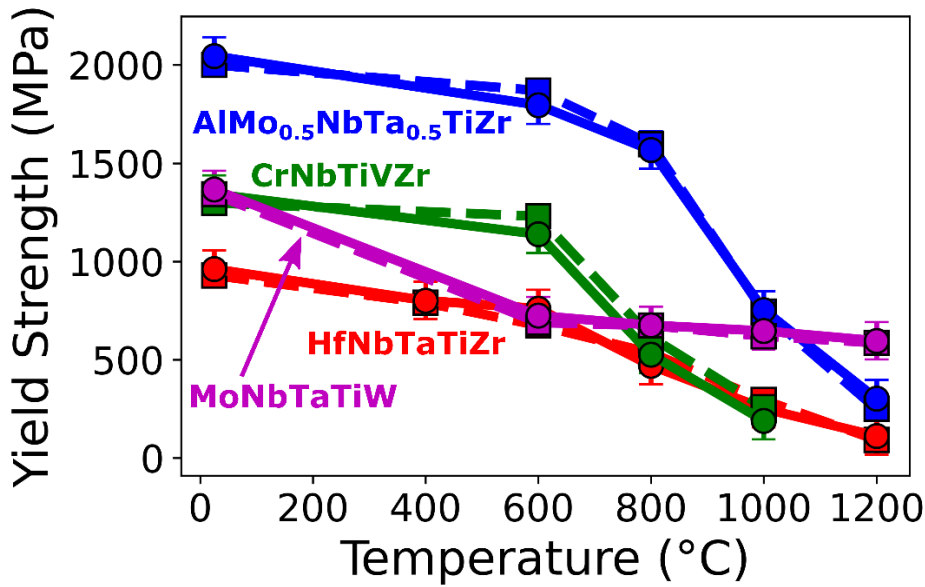

*Figure 3: Comparison of experimental (dashed lines, square symbols) and ML model predictions (solid lines, round symbols) of yield strength for four different RHEAs at various temperatures. Comparisons of all other RHEAs for which there are temperature-dependent experimental data are included in the SI.*

**Alloy Discovery: Known Base alloys as a Starting Point**

Once the ML model is developed, various approaches can be followed in order to discover new alloys with improved yield strength. The first approach, which experimentalists frequently follow, is to select a known alloy and improve its properties by varying the atomic fractions and/or adding new elements. The approach to our new RHEA discovery mimics this procedure computationally and is schematically shown in Figure 4a. The procedure varies the atomic composition as dictated by the genetic algorithm, and computes the yield strength using the forward model via computing the descriptors used in the model. Finally, convergence is achieved when the yield strength is maximized with respect to the elemental composition through the intelligent search of the compositional space towards the direction of improving yield strength. The details of the optimization process can be found in the Method section.

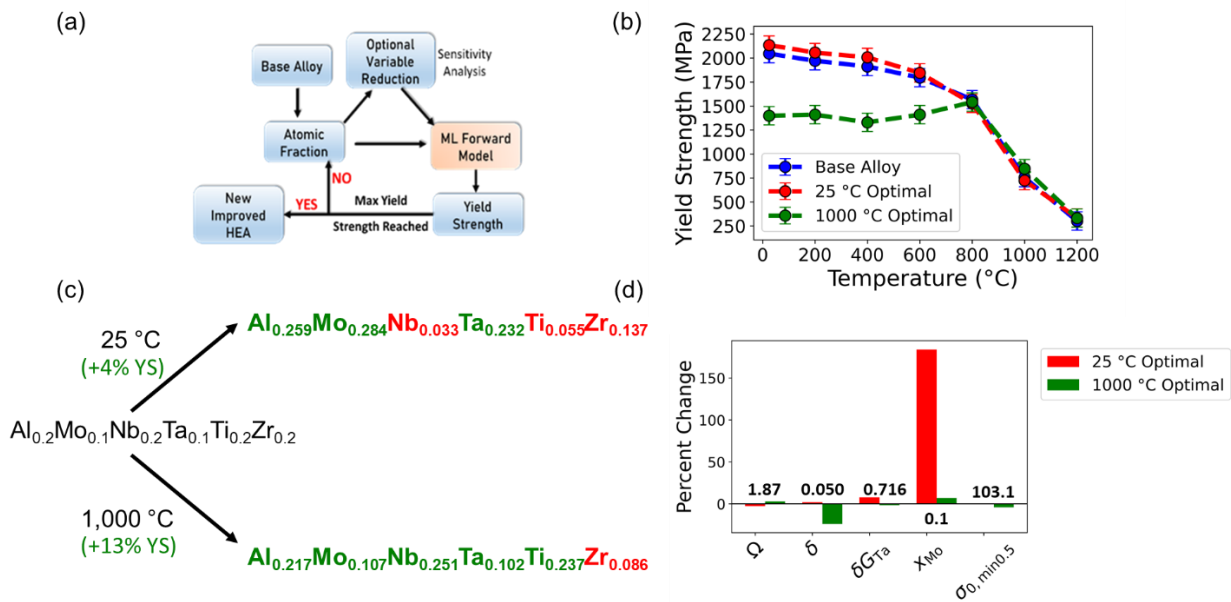

Figure 4: (a) The flowchart outlines the steps for optimizing the HEA composition from a base HEA. The only input required for the model is a base HEA composition; (b) yield strength vs. temperature profile for the original base alloy and the two optimized alloys. (c) comparison of the element fractions for the original base alloy and the alloys optimized for 25 °C and for 1,000 °C. Elements that were increased are shown in green text, while elements that were decreased are shown in red text; (d) percent change, relative to the base alloy, of each feature which serves as a direct input to the yield strength model. The numbers in bold black font are the values of each feature in the base alloy.

As a demonstration case, we first selected $AlMo_{0.5}NbTa_{0.5}TiZr$ which already shows high yield strength at high-temperature. Therefore, it is logical to use this alloy as a starting point for the genetic algorithm optimization to examine whether it is possible to improve its yield strength further by manipulating the composition. To begin the optimization, we chose a temperature at which to maximize the yield strength. As an example, we are primarily concerned with discovering RHEA compositions with maximized yield strength at room temperature (25 °C) and at elevated temperature (1,000 °C). The optimization progress was visualized by performing principal component analysis (Figures S7-S8). *Figure 4*b compares the yield strength vs. temperature profiles of the base composition, the compositions that maximizes yield strength at 25 °C and 1,000 °C. The base alloy and the 25 °C optimal alloy behave very similarly with respect to temperature. At 25 °C the optimized alloy improves upon the base alloy by 90 MPa, or 4%. The minor improvement in the yield strength indicates that the base alloy was already near optimal for yield strength at 25 °C. Furthermore, the base alloy and the 25 °C optimal alloy have a nearly identical temperature dependence. The 1,000 °C optimal alloy exhibits a notably different temperature dependence, with the yield strength being approximately constant between 25 – 800 °C. The 1,000 °C optimal alloy has a significantly reduced yield strength at 25 °C (1398 MPa), yet improves upon the base alloy yield strength by 13% at 1,000 °C (848 MPa). The remarkable temperature insensitivity of the 1,000 °C optimal alloy is particularly worth noting, since experimentalists have frequently concerned themselves with searching for alloys that maintain high yield strength as they are exposed to increasingly high temperatures. In this case, given that the value of $\delta$ is the prevailing difference between the 1,000 °C optimal alloy and the base alloy, we can conclude that lower values of $\delta$ are correlated to a lessened temperature dependence of the yield strength. In *Figure 4*c the element fractions of the base alloy (AlMo$_{0.5}$NbTa$_{0.5}$TiZr) is compared to the final composition of the alloys optimized for 25 °C and 1,000 °C yield strength. In the 25 °C alloy, the Mo fraction is shown to increase significantly relative to the base alloy. For the 1,000 °C optimized alloy, on the other hand, the Mo fraction remains approximately unchanged while the Zr fraction decreases significantly. Simulation results thus indicate composition that improves room temperature strength does not necessarily improves high-temperature yield strength. The approach developed here therefore particularly useful for predicting RHEA compositions customized to have superior yield strength at a specific temperature of interest. *Figure 4*d shows the changes for each of the descriptors serving as direct inputs to the ML model at two different temperatures. Evident from observing the changes in the descriptor values relative to the base alloy is that the $x_{Mo}$ constitutes by far the largest change for the 25 °C optimal alloy. On the other hand, $\delta$, the atomic size mismatch, undergoes the largest change for the 1,000 °C optimal alloy. That the optimal atomic size mismatch is significantly smaller for the 1,000 °C optimal alloy indicates that increasing lattice distortion could have a deleterious impact on the yield strength, particularly at high temperature. Our results suggest that the impact of model descriptors on the yield strength is complex and temperature-dependent, thereby pointing towards a different physics which should be considered when designing alloys for a particular temperature range.

The aforementioned inverse optimization for both 25 °C and 1,000 °C starting from $AlMo_{0.5}NbTa_{0.5}TiZr$ constitutes just one example of workflow that has been developed to improve upon a base alloy. The choice of the $AlMo_{0.5}NbTa_{0.5}TiZr$ RHEA to serve as the base alloy was informed by its already having an exceptional high-temperature yield strength. Other rationales could be used that could lead one to choose a different known alloy as a starting point for the optimization. For example, in addition to yield strength, ductility is a mechanical property of frequent concern. Yet, yield strength and ductility have a well-known tradeoff, where increasing ductility tends to lead to lower yield strength (Figure S9). Thus, it is logical to start with compositions that already possess "good" room temperature ductility and manipulate them to maximize yield strength through optimization. While there is no certainty that the new compositions with improved yield strengths will retain their ductility, ML models can be developed to maximize both yield strength and ductility simultaneously. Such models are currently under development by us where one can find compositions that simultaneously improve multiple properties. In this paper, we have chosen two additional base alloys, $Mo_{0.3}NbTiV_{0.3}Zr$ and HfNbTaTiZr (as shown in Figure 3), which show high room temperature ductility. $Mo_{0.3}NbTiV_{0.3}Zr$ has a moderate yield strength at 25 °C (1312 MPa) and a ductility of 49.3%[51]. HfNbTaTiZr also has a high ductility (33.3%) but a very poor YS at 25 °C (929 MPa)[49].

The optimizations were performed on these two base compositions at 25 °C and 1000 °C, and the results of their temperature dependent yield strengths are shown in *Figure 5*. Immediately apparent in *Figure 5* is that greater improvements in the yield strength are seen for these two cases relative to what was $AlMo_{0.5}NbTa_{0.5}TiZr$ case in *Figure 4*d. This is a consequence of both of these base alloys being particularly sub-optimal with respect to yield strength as they were selected for their good *ductility* instead. It can be seen that significant improvements were achieved at both 25 °C and 1000 °C over the base alloy compositions; in particular for HfNbTaTiZr, 80% improvement (increase from 962 MPa to 1731 MPa) was achieved at 25 °C, and 36% improvement (252 MPa to 344 MPa) was obtained at 1000 °C by optimization of the elemental composition.

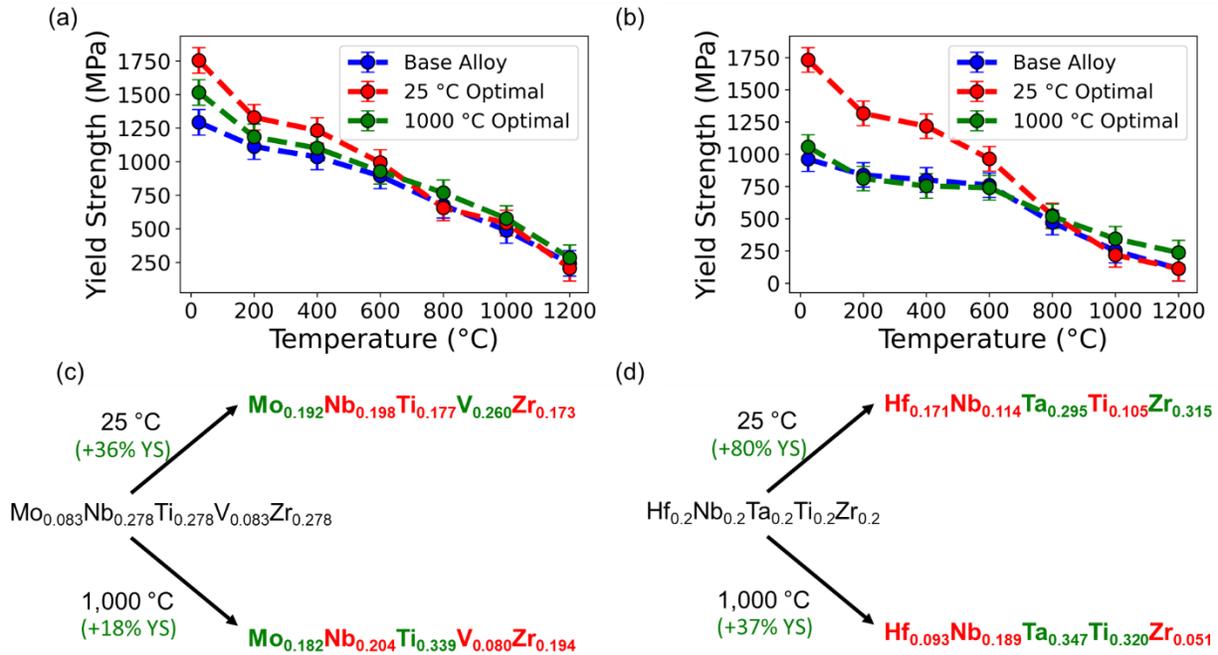

*Figure 5: Temperature-dependent yield strength predictions for the (a) $Mo_{0.3}NbTiV_{0.3}Zr$ base alloy case, and the (b) HfNbTaTiZr base alloy case. The original base alloy is shown together with the optimized alloy for 25 °C and the optimized alloy for 1,000 °C. The element compositions corresponding to each case are shown in panels (c) and (d). Element fractions that were increased during the optimization are shown in green text, while element fractions that were decreased are shown in red text.*

Inspecting the elemental compositions provided in *Figure 5* reveals some important trends with respect to which elements are modified to render an improved alloy. For the 25 °C optimal alloy in *Figure 5*a, the concentrations of Nb, Ti, and Zr were decreased by approximately equivalent amounts relative to the base alloy, whereas the concentration of Mo and V were both increased significantly. By contrast, for the 1,000 °C optimal alloy, the Ti fraction was increased and the V fraction was nearly unchanged. Similar to the alloy optimized for yield strength at 25 °C, however, the Mo concentration was also increased. That Mo generally tends to improved yield strength is consistent with the SHAP analysis discussed in Figure 2b. Likewise, considering the HfNbTaTiZr base alloy case in *Figure 5*b, Ti is also shown to be present in a high concentration for the 1,000 °C optimal alloy and in low concentration for the 25 °C optimal alloy. Meanwhile, Ta and Zr in the HfNbTaTiZr base alloy exhibit a similar relationship with the yield strength that Mo and V exhibited in the $Mo_{0.3}NbTiV_{0.3}Zr$ base alloy. That increasing the Ta element fraction often corresponds to improved yield strength was also evident from the earlier SHAP analysis where the $\delta G_{Ta}$ was shown to have a positive impact on the yield strength.

**Alloy Discovery: Inclusion of Additional Elements, and Novel Base Alloys Identification**

So far, we have restricted the compositional variation within the elements of a base alloy without adding or exchanging any elements. A total of 10 elements (i.e., Al, Cr, Hf, Mo, Nb, Ta, Ti, V, W, and Zr) are present in the training dataset. Inclusion of additional elements was first performed via using the equimolar 10-element alloy as a starting point. Comparison of the temperature-dependent yield strength for the 10-element base alloy, the 25 °C optimal, and 1,000 °C optimal is provided in *Figure 6*a. The comparison reveals that both optimized alloys result in a significant improvement at 25 °C, and in fact, the two optimized alloys have very similar yield strengths at 25 °C. However, at 1,000 °C, the 25 °C optimal alloy has a lower yield strength than the base alloy. The 1,000 °C optimal alloy, however, maintains a high yield strength resulting in a 48% improvement upon the base alloy. The elemental compositions are provided in *Figure 6*b. Common refractory elements (Hf, Nb, and Mo) are shown to be particularly beneficial when optimizing at 1,000 °C, whereas lighter elements (Al and Ti) are more beneficial at 25 °C.

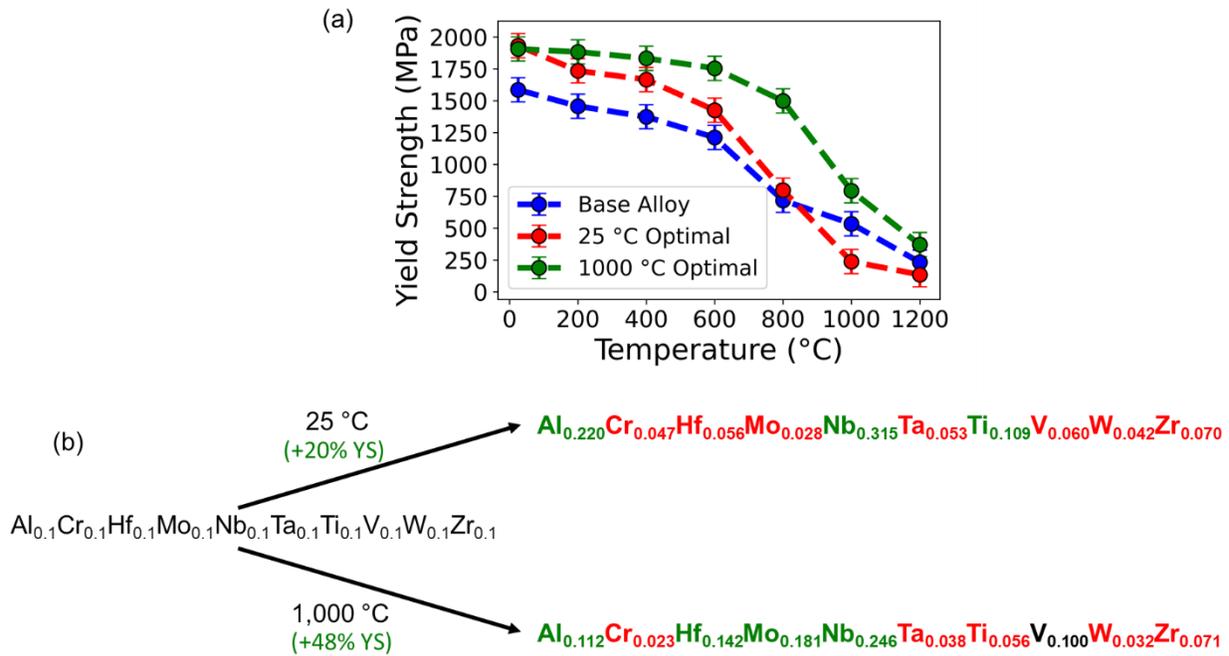

*Figure 6. (a) Temperature-dependent yield strength of the ten-element alloy. (b) Corresponding changes in elemental composition starting from the equimolar base alloy and optimizing at 25 °C and 1,000 °C.*

To mimic typical HEA compositions, which often include only 4 – 6 principal elements, we also extended our approach to systematically predict the yield strength for every unique five-element RHEA from the set of 10 elements. Choosing 5 elements from this set of 10 elements yields 252 unique equiatomic alloy combinations, some of which have been reported in the literature. The forward ML model was used to predict the yield strength of all 252 equiatomic alloys, at both 25 °C and 1,000 °C. The distribution of predicted yield strengths for 25 °C and 1,000 °C are shown in *Figure 7*a and *Figure 7*b, respectively. At 25 °C the yield strengths are approximately normally distributed with a mean of 1,457 MPa and a standard deviation of 175 MPa. The yield strength distribution at 1,000 °C, however, is clearly bimodal. One mode is located at ~200 MPa, while the other mode is located at ~550 MPa. The bimodal behavior at 1,000 °C is a result of the presence of Mo. If Mo is present in the alloy, the alloy belongs to the class of materials with an average predicted yield strength of ~550 MPa at 1,000 °C. In the absence of Mo, the alloy has a notably reduced high-temperature yield strength property.

The equiatomic alloys, AlMoTaTiZr and AlHfMoTaTi, with the highest yield strengths at 25 °C and at 1,000 °C, respectively, were identified and selected for further optimization. The two alloys are notably similar, with the only difference being the replacement of Zr with Hf. Both of these equiatomic alloys were not present in the compiled Couzinie *et al.* RHEA dataset[34] used for training and validation, nor, to the best of our knowledge, have they been reported in any of the RHEA literature. In *Figure 7*c and *Figure 7*d, the yield strengths of these two novel equiatomic alloys were further improved through optimization of their element fractions. Since both equiatomic base alloys had high yield strength, optimizing the element fractions resulted in only slight improvements to the yield strength, with the $Al_{0.239}Mo_{0.123}Ta_{0.095}Ti_{0.342}Zr_{0.201}$ alloy resulting in an improvement of 7% at 25 °C, and the $Al_{0.151}Hf_{0.236}Mo_{0.137}Ta_{0.131}Ti_{0.345}$ alloy resulting in an improvement of 2% at 1,000 °C. Also, worth noting is that, assuming a rule-of-mixtures, the $Al_{0.239}Mo_{0.123}Ta_{0.095}Ti_{0.342}Zr_{0.201}$ alloy had a density of only 6.3 g cm$^{-3}$. According to the Couzinie *et al.* dataset[34], there is no HEA which is both stronger at 25 °C and less dense than the $Al_{0.239}Mo_{0.123}Ta_{0.095}Ti_{0.342}Zr_{0.201}$ RHEA.

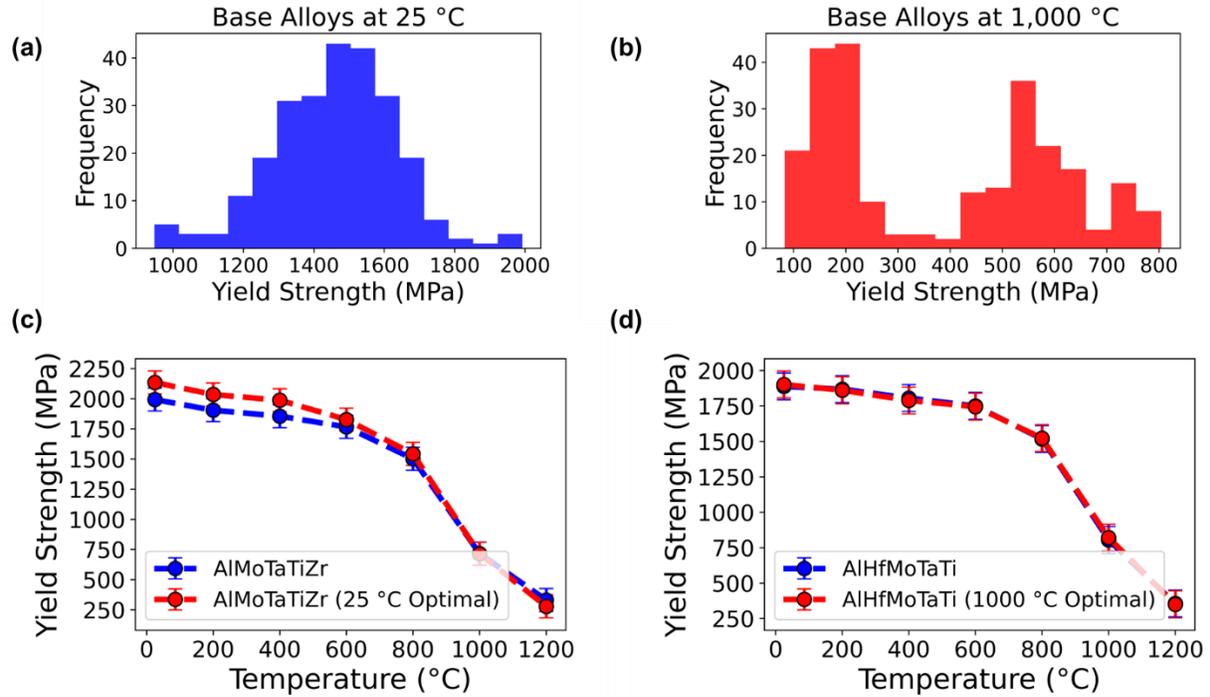

*Figure 7: Yield strength predictions of 252 unique five-element equiatomic base alloys. (a) Distribution of yield strength at 25 °C for the five-element equiatomic alloys. (b) Distribution of yields strength at 1,000 °C. (c) AlMoTaTiZr and $Al_{0.239}Mo_{0.123}Ta_{0.095}Ti_{0.342}Zr_{0.201}$ (optimized for 25 °C). Yield strength improved by 7%. (d) AlHfMoTaTi and $Al_{0.151}Hf_{0.236}Mo_{0.137}Ta_{0.131}Ti_{0.345}$ (optimized for 1,000 °C). Yield strength improved by 2%.*

**Conclusions**

In this paper, we have demonstrated the concept of an intelligent computational framework based on machine learning and optimization to predict RHEA yield strength and discover new compositions with theoretical improvement over the starting RHEA. The protocol developed here can be used by material scientists for quick screening of compositional space and identifying potential candidates with improved yield strength that merit processing and characterization. First, we have shown that repeated *k*-fold cross-validation coupled with feature selection is an effective approach to obtain a more statistically meaningful prediction of all data points in contrast to traditional used ML validation techniques. Using the robust ML-based yield strength prediction model with a clear understanding of the statistical errors, we have coupled this with a genetic algorithm to discover new RHEA compositions with improved yield strengths. Given a baseline starting RHEA, the algorithm intelligently searches through the complex composition space to maximize yield strength. The concept was demonstrated for three different base alloys discussed in the RHEA literature, and optimal alloy compositions with improved yield strengths, as high as 80% were predicted for both 25 °C and 1,000 °C. The alloys optimized for yield strength at 25 °C and at 1,000 °C exhibited notable differences in composition and descriptors, underscoring that the mechanisms and criteria for maximizing strength at low temperature and high temperature can be quite different, and compositions maximizing yield strength at room temperature may not improve that for high temperatures. Finally, in a generalized approach, we predicted the low-temperature and high-temperature yield strength of 252 equiatomic RHEA chemistries. The top candidate for each temperature was further improved by tailoring the elemental composition using our generalized framework. Our ongoing work is extending this technique to predict other mechanical properties, such as hardness, ductility/plasticity, creep strength, and fatigue. The simultaneous optimization of multiple properties can also be incorporated into the framework. The ability to perform multi-property optimization will enable discovering new HEAs with, for example, high ductility at room temperature and high strength at higher temperatures. Identifying HEAs which meet requirements for multiple properties experimentally is a grand challenge, and the work presented here creates a foundation for addressing this challenge.

**Methods**

Some noteworthy descriptors that bear discussion are ones that have been traditionally used for interpreting phases, such as the atomic size mismatch ($\delta$), the enthalpy of mixing ($\Delta H_{\text{mix}}$), the entropy of mixing ($\Delta S_{\text{mix}}$), $\Omega$, and $\Phi$. The equations defining these five descriptors are given as[17,18],

$$\delta = \sqrt{\sum_{i=1}^{n} x_i(1 - r_i/\bar{r})^2} \quad (1)$$

$$\Delta H_{\text{mix}} = \sum_{i=1, i \neq j}^{n} \Omega_{ij} x_i x_j \quad (2)$$

$$\Delta S_{\text{mix}} = -R \sum_{i=1, i \neq j}^{n} x_i \ln x_j \quad (3)$$

$$\Omega = \frac{T_m \Delta S_{\text{mix}}}{|\Delta H_{\text{mix}}|} \quad (4)$$

$$\Phi = \left|\frac{\Delta H_{\text{mix}} - T_m \Delta S_{\text{mix}}}{\Delta H_{\text{max}} - T_{\text{max}} \Delta S_{\text{max}}}\right| \quad (5)$$

In Equations 1 – 5, $i$ and $j$ represent the $i^{\text{th}}$ and $j^{\text{th}}$ element, $r_i$ is the atomic radius of the $i^{\text{th}}$ element, $\bar{r}$ is the average atomic radius of the alloy, $\Omega_{ij} = 4\Delta H_{\text{AB}}^{\text{mix}}$ (where $\Delta H_{\text{AB}}^{\text{mix}}$ is the enthalpy of mixing of binary alloys), $x$ is the element fraction, $T_m$ is melting temperature estimated from a rule-of-mixtures, $\Delta H_{\text{max}}$ is the maximum enthalpy of mixing of a binary combination of the elements in the alloy, $T_{\text{max}}$ is the maximum melting temperature of a single element in the alloy, and $\Delta S_{\text{max}}$ is the maximum entropy of mixing of a binary alloy (i.e., an equiatomic binary alloy).

Recently, lattice distortion and moduli of distortion are thought to be important parameters for determining phase formation and mechanical properties of HEAs[21,52]. Therefore, we have included them in our descriptor calculation. The lattice distortion around atom $i$ can be defined according to Senkov et al.[48] as,

$$\delta_{ai} = \frac{9}{8} \sum_{j} x_j \delta_{aij} \quad (6)$$

The 9 in the numerator is due to total number of atoms in the $i$-centered cluster in the BCC lattice, while the 8 in the denominator is due to the number of atoms around $i$ in the cluster (excluding $i$). The reduced atomic size difference, $\delta_{aij}$, is defined as,

$$\delta_{aij} = \frac{2(r_i - r_j)}{(r_i + r_j)} \tag{7}$$

Similarly, the modulus of distortion, $\delta_{Gi}$, is the defined as,

$$\delta_{Gi} = \frac{9}{8}\sum_j c_j \delta_{Gij} \tag{8}$$

$$\delta_{Gij} = \frac{2(G_i - G_j)}{(G_i + G_j)} \tag{9}$$

where $G$ is the shear modulus. The atomic radii and moduli were collected from published values[17,53].

In addition to solid solution strengthening, grain boundary strengthening is another mechanism through which the mechanical properties of an alloy can be affected. The primary empirical equation governing the contribution of grain boundary strengthening to the observed mechanical property is the Hall-Petch relationship[54,55]. The Hall-Petch relationship is frequently expressed as,

$$\sigma_{\text{HP}} = \sigma_0 + Kd^{-\frac{1}{2}} \tag{10}$$

Where $\sigma_0$ is the base strength of the material, $K$ is the locking parameter, and $d$ is the grain size of the material. In alloys, the base strength is typically derived from the the individual element yield strengths using the rule-of-mixtures. The locking parameters of all elements of interest in this study have been tabulated in the literature[24]. Here, a rule-of-mixtures has been used to estimate the locking parameter of each alloy based on their elemental composition. The grain size, $d$, is closely related to the specific processing conditions affecting grain growth kinetics[50,56], and is typically not reported in the HEA literature. However, some HEA literature have performed detailed studies on the effect of ubiquitously reported processing conditions (e.g., annealing temperature, annealing time, etc.) on the grain size. Therefore, the specific processing conditions which we have collected from the original literature cited by Couzinie *et al.*[34] serve as a surrogate for a more detailed microstructural knowledge.

Finally, we have utilized the temperature-dependent yield strength model formulated by Varvenne and Curtin[44] to augment the experimentally available HEA dataset at temperatures between 25 °C and 600 °C. The accuracy of ML approaches is always fundamentally limited by the availability of adequate data, and RHEA yield strengths at temperatures between 25 °C and 600 °C are typically not reported. Using the experimental yield strengths at 25 °C and 600 °C and the temperature-dependent yield strength model proposed by Varvenne and Curtin[44], the yield strength for 200 °C and 400 °C were computed. More details and an example calculation of the yield strength at 200 °C and 400 °C based on experimentally known values at 25 °C and 600 °C are provided in the SI.

**Descriptor Selection, Model Training and Validation**

Descriptor selection constitutes an important part of our work. *Figure 8*a outlines the procedure for training and validation of the ML model. We used the sequential forward selection (SFS) method[37,38] for selecting the best set of descriptors that describes the data within a given regression model. We have investigated four different regression models, namely: random forest[36], gradient boosting[41], LASSO[39], and ridge regression[40]. The procedure starts with first performing SFS coupled with a *k*-fold cross-validation technique where *k* is taken as five (5). In 5-fold cross-validation, the data were randomly divided into five sets, with four sets are used for training and the rest is for testing. The 5-fold cross-validation was repeated until all groups are used for validation. During this process, a set of best descriptors were identified with a specified number of maximum descriptors criteria. We have systematically varied the number of descriptors in the computed 5-fold validation to identify a critical number of descriptors beyond which there was negligible improvement of the cross-validation regression coefficient $R^2$ (CV). The cross-validated $R^2$, $R^2$ (CV), remains constant beyond a six-descriptor model (*Figure 8*b) which indicates that there is insignificant improvement in the predicted yield strength if more than six descriptors are chosen. We then performed optimization of the hyperparameters of the regression models using the six descriptors to identify parameters that maximize $R^2$ (CV). Once we have identified the descriptor set and optimized the hyperparameters, we performed a repeated *k*-fold validation. In the procedure, we performed the repeated 5-fold cross-validation step 1,000 times. Each time, the members of each fold were chosen randomly which eliminated the biasness of a single 5-fold analysis. The repeated *k*-fold method provides a statistical variation of the predicted yield strength of *each* RHEA allowing us to compute the 95% confidence level of *each* yield strength data point. We believe that our procedure provides a superior assessment of model's predictive ability compared to the commonly used single *k*-fold validation, or train-test splitting. At the end, we also performed 25 runs of SFS with six descriptors in order to examine variability of the selected features. The resulting feature selection probability distribution is provided in Figure S10. Replicates of the SFS process revealed that multiple composition-based descriptors had an approximately equal likelihood of being selected. Since multiple feature sets yielded nearly the same predictions, a representative feature set was chosen for the forward model and discovery process.

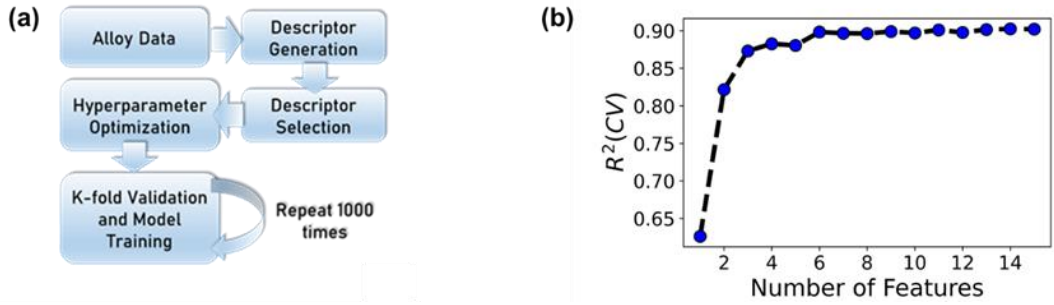

*Figure 8. (a) ML model development procedure followed in this work; (b) Changes in cross-validation coefficient as a function of number of descriptors. It shows there is no significant improvement beyond six (6) descriptors.*

**Yield Strength Optimization**

For solving the inverse problem, we used the differential evolution optimizer[33], a genetic algorithm, to design an atomic composition that maximizes yield strength at a specified temperature for a particular base alloy. Differential evolution is a stochastic population-based method that is frequently used for global optimization problems. At each pass through the population, the algorithm mutates each candidate solution by mixing with other candidate solutions to create a trial candidate. A central feature of the optimization problem is the definition of the *objective function*. For a minimization problem, the goal is to identify a solution that causes the objective function to be equal to zero, or minimizes the objective function to be as close to zero as possible. In this case, the goal is to maximize the yield strength. Therefore, the objective function should grow smaller as the yield strength grows larger. In the current investigation, we have found that defining the objective function as the reciprocal of the yield strength allows for alloys with increased yield strength to be found easily. The mutation constant was set between 0.5 and 1, with dithering employed. The recombination constant (i.e., crossover probability) was 0.7. A linear constraint function was coupled with the optimization using Lampinen's approach[57] to ensure that the element fractions summed to 1. All individual element fractions that were present in a given alloy were bounded between 0.02 and 0.35, typical values that are representative of an HEA.

**Acknowledgments**

We would like to thank Prof. Peter Liaw for helpful discussions. This work was funded by the Office of Naval Research of the United States under the Small Business Technology Transfer program (contract # N68335-20-C-0402).

**Data Availability**

The data used to generate the machine learning models (compositions, feature values, measured yield strengths, etc.) are included in the SI.

**Code Availability**

Readers are requested to contact the authors.

**References**

1. Yeh, J.-W. *et al.* Nanostructured High-Entropy Alloys with Multiple Principal Elements: Novel Alloy Design Concepts and Outcomes. *Adv. Eng. Mater.* **6**, 299–303 (2004).

2. Chen, T. K., Shun, T. T., Yeh, J. W. & Wong, M. S. Nanostructured nitride films of multi-element high-entropy alloys by reactive DC sputtering. *Surf. Coatings Technol.* **188**–**189**, 193–200 (2004).

3. Hsu, C.-Y., Yeh, J.-W., Chen, S.-K. & Shun, T.-T. Wear resistance and high-temperature compression strength of Fcc CuCoNiCrAl0.5Fe alloy with boron addition. *Metall. Mater. Trans. A* **35**, 1465–1469 (2004).

4. Huang, P.-K., Yeh, J.-W., Shun, T.-T. & Chen, S.-K. Multi-Principal-Element Alloys with Improved Oxidation and Wear Resistance for Thermal Spray Coating. *Adv. Eng. Mater.* **6**, 74–78 (2004).

5. Yeh, J.-W. *et al.* Formation of simple crystal structures in Cu-Co-Ni-Cr-Al-Fe-Ti-V alloys with multiprincipal metallic elements. *Metall. Mater. Trans. A* **35**, 2533–2536 (2004).

6. Cantor, B., Chang, I. T. H., Knight, P. & Vincent, A. J. B. Microstructural development in equiatomic multicomponent alloys. *Mater. Sci. Eng. A* **375**–**377**, 213–218 (2004).

7. Ranganathan, S. Alloyed pleasures: Multimetallic cocktails. *Curr. Sci.* **85**, 1404–1406 (2003).

8. Yeh, J.-W. Overview of High-Entropy Alloys. in *High-Entropy Alloys* (eds. Gao, M. C., Yeh, J.-W., Liaw, P. K. & Zhang, Y.) 1–19 (Springer, 2016).

9. Senkov, O. N., Senkova, S. V., Miracle, D. B. & Woodward, C. Mechanical properties of

low-density, refractory multi-principal element alloys of the Cr-Nb-Ti-V-Zr system. *Mater. Sci. Eng. A* **565**, 51–62 (2013).

10. Kang, B., Lee, J., Ryu, H. J. & Hong, S. H. Ultra-high strength WNbMoTaV high-entropy alloys with fine grain structure fabricated by powder metallurgical process. *Mater. Sci. Eng. A* **712**, 616–624 (2018).

11. Liu, Y. *et al.* Microstructure and mechanical properties of refractory HfMo0.5NbTiV0.5Sixhigh-entropy composites. *J. Alloys Compd.* **694**, 869–876 (2017).

12. Maiti, S. & Steurer, W. Structural-disorder and its effect on mechanical properties in single-phase TaNbHfZr high-entropy alloy. *Acta Mater.* **106**, 87–97 (2016).

13. Senkov, O. N., Isheim, D., Seidman, D. N. & Pilchak, A. L. Development of a refractory high entropy superalloy. *Entropy* **18**, (2016).

14. Zhang, M., Zhou, X. & Li, J. Microstructure and Mechanical Properties of a Refractory CoCrMoNbTi High-Entropy Alloy. *J. Mater. Eng. Perform.* **26**, 3657–3665 (2017).

15. Zhang, Y., Yang, X. & Liaw, P. K. Alloy design and properties optimization of high-entropy alloys. *JOM* vol. 64 830–838 (2012).

16. Miracle, D. B. High entropy alloys as a bold step forward in alloy development. *Nat. Commun.* **10**, 1–3 (2019).

17. Miracle, D. B. & Senkov, O. N. A critical review of high entropy alloys and related concepts. *Acta Mater.* **122**, 448–511 (2017).

18. Senkov, O. N., Miracle, D. B., Chaput, K. J. & Couzinie, J.-P. Development and exploration of refractory high entropy alloys—A review. *J. Mater. Res.* **33**, 3092–3128 (2018).

19. Yang, X. & Zhang, Y. Prediction of high-entropy stabilized solid-solution in multi-component alloys. *Mater. Chem. Phys.* **132**, 233–238 (2012).

20. Chen, H. *et al.* Contribution of Lattice Distortion to Solid Solution Strengthening in a Series of Refractory High Entropy Alloys. *Metall. Mater. Trans. A Phys. Metall. Mater. Sci.* **49**, 772–781 (2018).

21. Coury, F. G., Clarke, K. D., Kiminami, C. S., Kaufman, M. J. & Clarke, A. J. High Throughput Discovery and Design of Strong Multicomponent Metallic Solid Solutions. *Sci. Rep.* 1–10 (2018) doi:10.1038/s41598-018-26830-6.

22. Yao, H. W. *et al.* Mechanical properties of refractory high-entropy alloys: Experiments and modeling. *J. Alloys Compd.* **696**, 1139–1150 (2017).

23. Wang, Z., Fang, Q., Li, J., Liu, B. & Liu, Y. Effect of lattice distortion on solid solution strengthening of BCC high-entropy alloys. *J. Mater. Sci. Technol.* (2018) doi:10.1016/j.jmst.2017.07.013.

24. Cordero, Z. & Schuh, C. A. Six decades of the Hall – Petch effect – a survey of grain-size strengthening studies on pure metals. (2016) doi:10.1080/09506608.2016.1191808.

25. Asghari-Rad, P. *et al*. Effect of grain size on the tensile behavior of V10Cr15Mn5Fe35Co10Ni25 high entropy alloy. *Mater. Sci. Eng. A* **744**, 610–617 (2019).

26. Schuh, B. *et al*. Thermodynamic instability of a nanocrystalline, single-phase TiZrNbHfTa alloy and its impact on the mechanical properties. *Acta Mater.* **142**, 201–212 (2018).

27. Tang, Z., Zhang, S., Cai, R., Zhou, Q. & Wang, H. Designing High Entropy Alloys with Dual fcc and bcc Solid-Solution Phases : Structures and Mechanical Properties. *Metall. Mater. Trans. A* **50**, 1888–1901 (2019).

28. He, J. Y. *et al. A precipitation-hardened high-entropy alloy with outstanding tensile properties*. http://www.elsevier.com/open-access/userlicense/1.0/ (2015).

29. Maresca, F. *et al*. Edge Dislocations Can Control Yield Strength in Refractory Body-Centered-Cubic High Entropy Alloys. *arXiv* 2008.11671 (2020).

30. Lee, S. Y., Byeon, S., Kim, H. S., Jin, H. & Lee, S. Deep learning-based phase prediction of high-entropy alloys: Optimization, generation, and explanation. *Mater. Des.* **197**, 109260 (2021).

31. Huang, W., Martin, P. & Zhuang, H. L. Machine-learning phase prediction of high-entropy alloys. *Acta Mater.* **169**, 225–236 (2019).

32. Zhou, Z., Zhou, Y., He, Q., Ding, Z. & Li, F. Machine learning guided appraisal and exploration of phase design for high entropy alloys. *npj Comput. Mater.* **5**, 128.

33. Storn, R. & Price, K. Differential Evolution - a Simple and Efficient Heuristic for Global Optimization over Continuous Spaces. *J. Glob. Optim.* **11**, 341–359 (1997).

34. Couzinié, J. P., Senkov, O. N., Miracle, D. B. & Dirras, G. Comprehensive data compilation on the mechanical properties of refractory high-entropy alloys. *Data Br.* **21**, 1622–1641 (2018).

35. Ward, L. *et al*. Matminer: An open source toolkit for materials data mining. *Comput. Mater. Sci.* **152**, 60–69 (2018).

36. Breiman, L. Random forests. *Mach. Learn.* **45**, 5–32 (2001).

37. Pudil, P., Novoviĉová, J. & Kittler, J. Floating search methods in feature selection. *Pattern Recognit. Lett.* **15**, 1119–1125 (1994).

38. Ferri, F. J., Pudil, P., Hatef, M. & Kittler, J. Comparative study of techniques for large-scale feature selection. *Pattern Recognit. Pract. IV* 403–413.

39. Tibshirani, R. Regression Shrinkage and Selection via the Lasso. *J. R. Stat. Soc. B* **58**, 267–288 (1996).

40. Hoerl, A. E. & Kennard, R. W. Ridge Regression: Applications to Nonorthogonal Problems. *Technometrics* **12**, 69–82 (1970).

41. Friedman, J. H. Greedy function approximation: A gradient boosting machine. *Ann. Stat.* **29**, 1189–1232 (2001).

42. Lundberg, S. M. & Lee, S. I. A unified approach to interpreting model predictions. *31st*

*Conf. Neural Inf. Process. Syst. (NIPS 2017)* 4766–4775 (2017).

43. Lundberg, S. M. *et al.* From local explanations to global understanding with explainable AI for trees. *Nat. Mach. Intell.* **2**, 56–67 (2020).

44. Varvenne, C., Luque, A. & Curtin, W. A. Theory of strengthening in fcc high entropy alloys. *Acta Mater.* **118**, 164–176 (2016).

45. Senkov, O. N., Jensen, J. K., Pilchak, A. L., Miracle, D. B. & Fraser, H. L. Compositional variation effects on the microstructure and properties of a refractory high-entropy superalloy AlMo0.5NbTa0.5TiZr. *Mater. Des.* **139**, 498–511 (2018).

46. Senkov, O. N., Woodward, C. & Miracle, D. B. Microstructure and Properties of Aluminum-Containing Refractory High-Entropy Alloys. *JOM* **66**, 2030–2042 (2014).

47. Senkov, O. N., Senkova, S. V. & Woodward, C. Effect of aluminum on the microstructure and properties of two refractory high-entropy alloys. *Acta Mater.* **68**, 214–228 (2014).

48. Senkov, O. N., Scott, J. M., Senkova, S. V., Miracle, D. B. & Woodward, C. F. Microstructure and room temperature properties of a high-entropy TaNbHfZrTi alloy. *J. Alloys Compd.* **509**, 6043–6048 (2011).

49. Senkov, O. N. *et al.* Microstructure and elevated temperature properties of a refractory TaNbHfZrTi alloy. *J. Mater. Sci.* **47**, 4062–4074 (2012).

50. Juan, C. C. *et al.* Simultaneously increasing the strength and ductility of a refractory high-entropy alloy via grain refining. *Mater. Lett.* **184**, 200–203 (2016).

51. Wu, Y. D. *et al.* Phase composition and solid solution strengthening effect in TiZrNbMoV high-entropy alloys. *Mater. Des.* **83**, 651–660 (2015).

52. Song, H. *et al.* Local lattice distortion in high-entropy alloys. *Phys. Rev. Mater.* **1**, 023404 (2017).

53. AZO Materials. https://www.azom.com/.

54. Hall, E. O. The Deformation and Ageing of Mild Steel: III Discussion of Results. *Proc. Phys. Soc. Sect. B* **64**, 747 (1951).

55. Petch, N. J. The Cleavage Strength of Polycrystals. *J. Iron Steel Inst.* **174**, 25–28 (1953).

56. Fazakas, E. *et al.* Experimental and theoretical study of Ti20Zr20Hf 20Nb20X20 (X = v or Cr) refractory high-entropy alloys. *Int. J. Refract. Met. Hard Mater.* **47**, 131–138 (2014).

57. Lampinen, J. A constraint handling approach for the differential evolution algorithm. *Proc. 2002 Congr. Evol. Comput. CEC 2002* **2**, 1468–1473 (2002).

# SUPPORTING INFORMATION

## Alternative Predicted Models

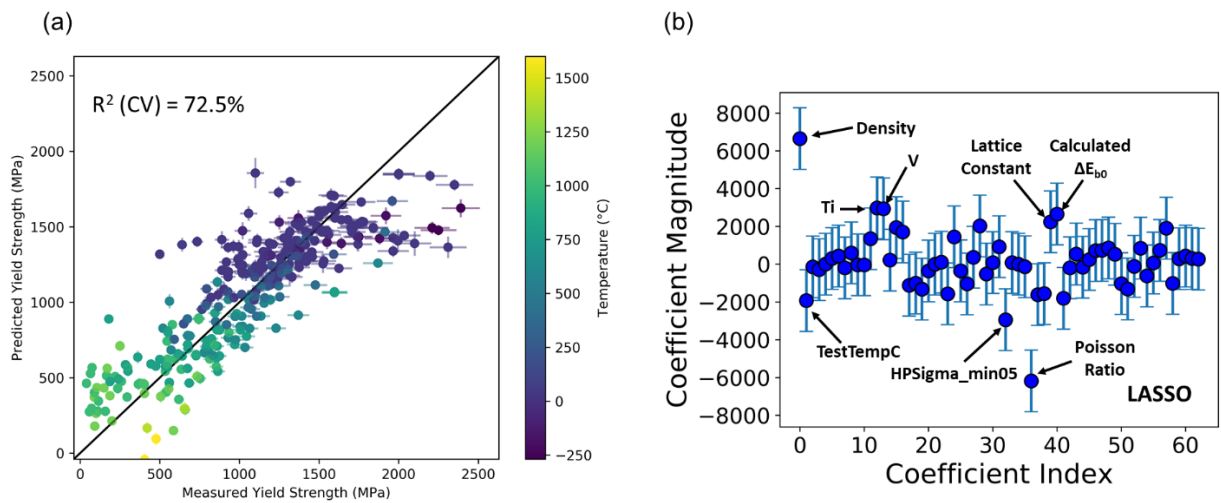

*Figure S1. LASSO regression: (a) parity plot comparison to experimental values; (b) descriptor coefficients following L1 regularization. Most important descriptors have been annotated.*

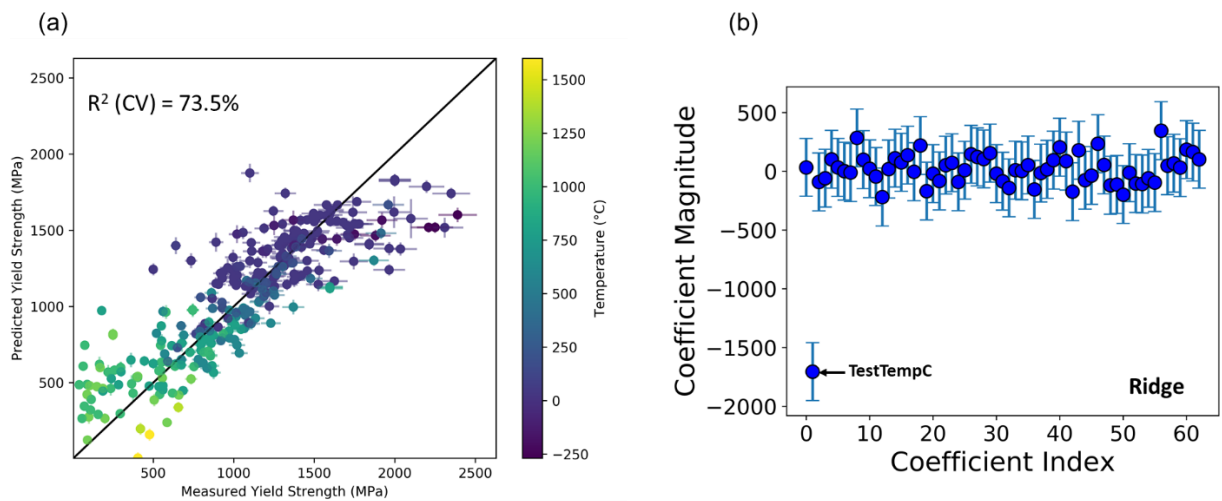

*Figure S2. Ridge regression: (a) parity plot comparison to experimental values; (b) descriptor coefficients following L2 regularization. The test temperature is the most important descriptor by far.*

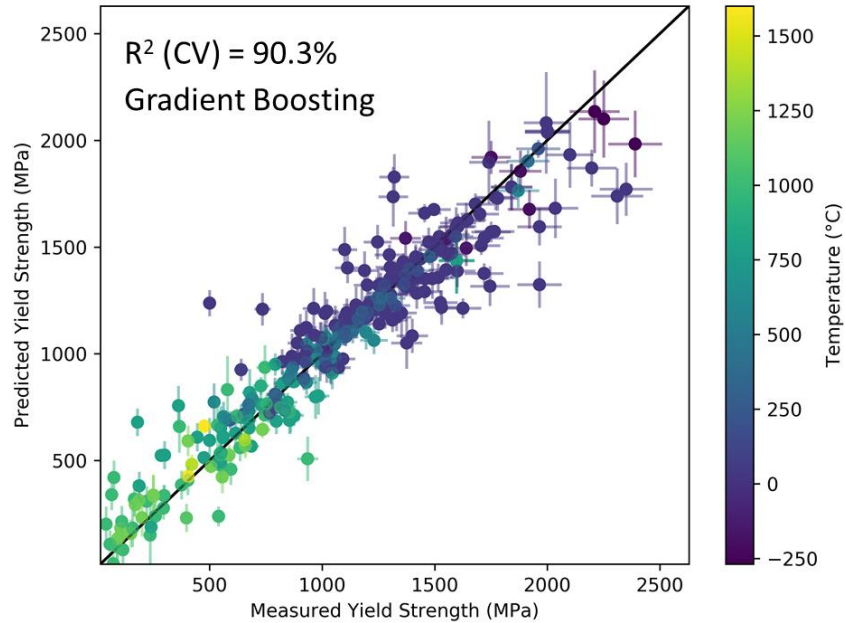

*Figure S3. Parity plot comparison for the gradient boosting model. The $R^2$ (CV) is approximately equivalent to what was obtained for the random forest model.*

**Dependence of Yield Strength on Ω from SHAP Analysis**

As was noted in the discussion alongside **Error! Reference source not found.** in regards to the SHAP explainability analysis that was performed for the forward model, Ω has a complex, but important effect on the predicted yield strength. As shown in Figure S4, small values of Ω result in positive SHAP values (i.e., an increase in the predicted yield strength), whereas large values of Ω result in a negative impact on the yield strength. Notably, this shift in the effect of Ω undergoes a transition across a small regime. It is theorized that this could result from the interplay of the entropy and enthalpy of mixing. Large values of Ω result when the enthalpy of mixing approaches zero, which may be detrimental to improving yield strength.

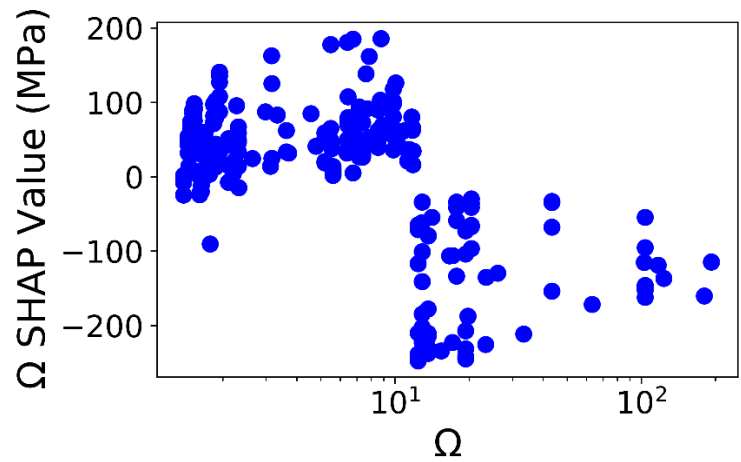

*Figure S4. Dependence of the SHAP value of Ω on the value of Ω.*

**Percentage Error Distribution of Our Model**

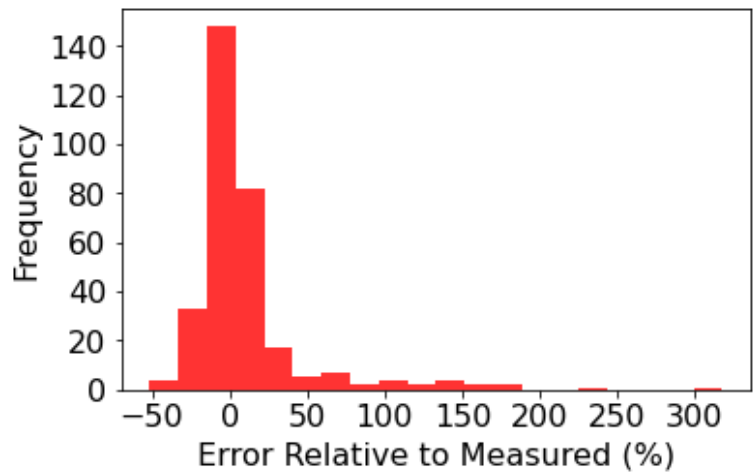

*Figure S5. Percentage error distribution of the model presented in this work relative to the measured values. MAE = 20.1%.*

**Comparison of Error in Analytical Model by Varvenne and Curtin**

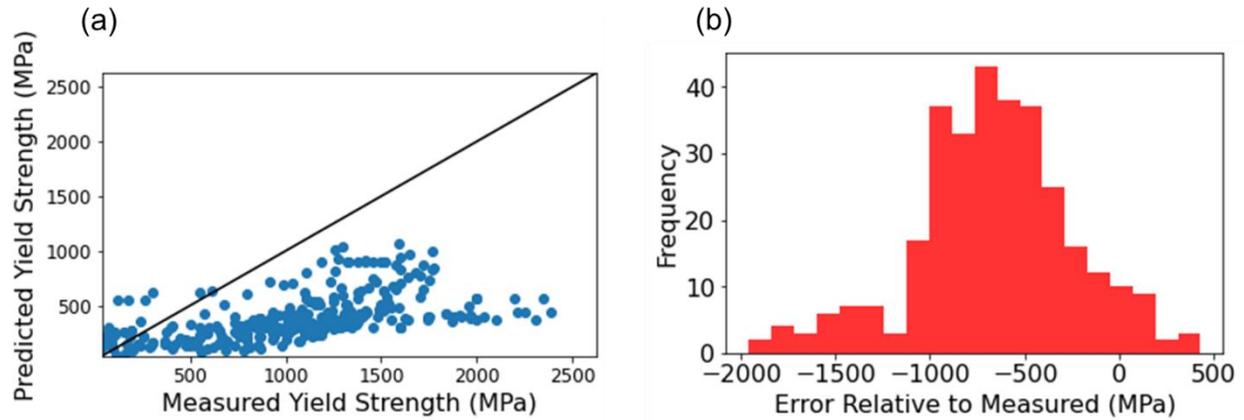

*Figure S6. Predictions obtained using the analytical temperature-dependent yield strength model proposed by Varvenne and Curtin. (a) parity plot of measured values and predicted values from the Varvenne and Curtin model. (b) distribution of error relative to measured values for the Varvenne and Curtin model. MAE = 683 MPa.*

**Interpretation Using Principal Component Analysis**

As mentioned above, the YS model provided here employs six descriptors: test temperature, $\Omega$, atomic size mismatch ($\delta$), modulus distortion of tantalum ($\delta G_{Ta}$), molybdenum atomic fraction, $x_{Mo}$, and a base strength quantity, $\sigma_{0, min0.5}$. Given that the input to the model has been limited to only these six descriptors, an unsupervised approach such as principal component analysis (PCA) can be used to further reduce the dimensionality and allow for clustering of the data in the descriptor space to be visualized.

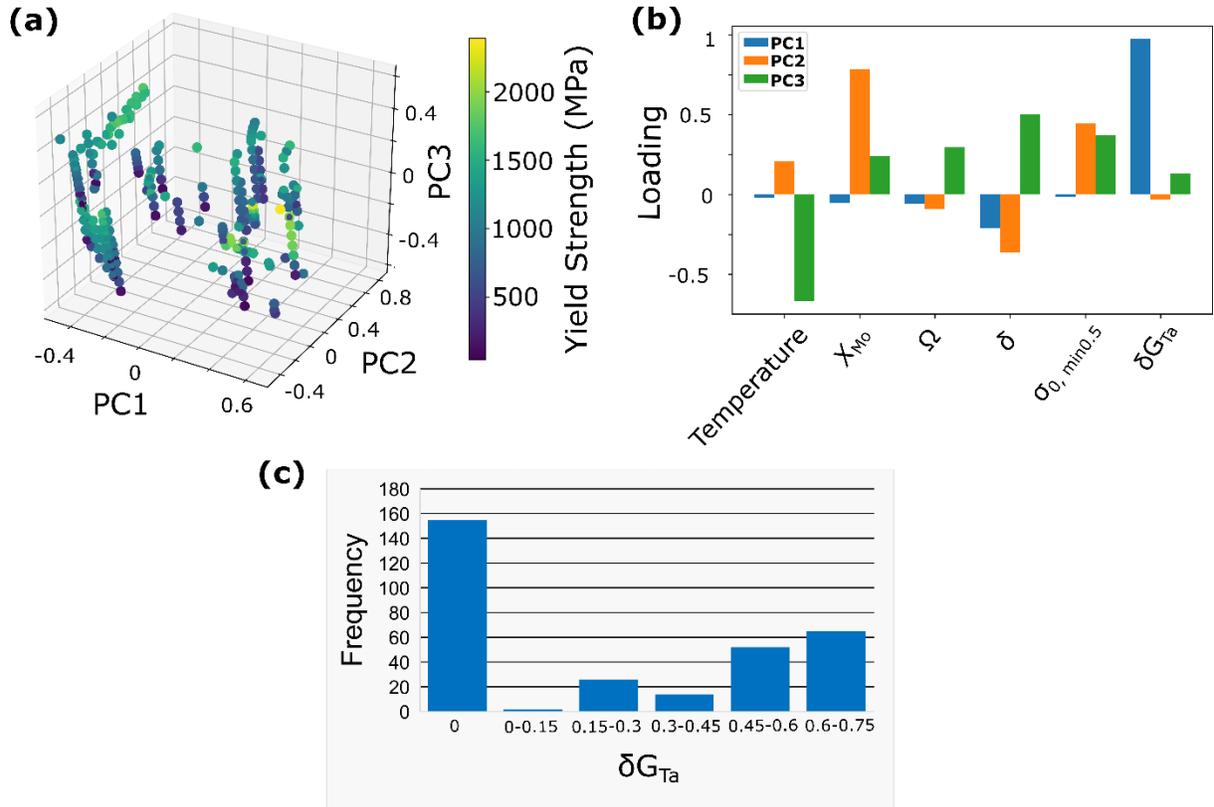

*Figure S7: (a) Three-dimensional score plot of compression yield strength data according to their first, second, and third principal component values; (b) Loadings of the six original descriptors on the first three principal components; (c) Histogram of tantalum modulus distortion values. The 155 alloys with δG$_{Ta}$ equal to zero correspond to non-tantalum-containing alloys.*

Analysis of the corresponding loading coefficients of the original six descriptors on the three principal components reveals the origin of the clustering observed in Figure S7a. The loadings are provided in Figure S7b. The loadings reveal that the first principal component is primarily composed of, and positively correlated to, the tantalum modulus distortion (δG$_{Ta}$). As a result, the clustering observed in Figure S7 (a) is in fact distinguishing between Ta and non-Ta HEAs. The Couzinie data[1] is approximately balanced with regards to Ta alloys, with 159 alloys (50.6%) containing Ta, and 154 alloys (49.4%) not containing Ta. The data clustering in Figure S7a based on Ta content is further evident upon analyzing the effect of Ta on modulus distortion. The Ta modulus distortion values for the RHEA dataset is provided in Figure S7(c). As can be seen, in the alloys which do not contain Ta, $\delta G_{Ta}$ is equal to zero. When Ta is present in the alloy, however, $\delta G_{Ta}$ always has a positive value. This is due to Ta having a high elemental shear modulus (69 GPa) relative to most of the other elements of interest. From a physical point of view, a positive modulus distortion implies that the presence of Ta usually leads to an increase in the shear modulus of the alloy. Another conclusion which can be drawn from the loading plot is that an increase in test temperature is primarily responsible for decreasing PC3 which has negative effects on the yield strength (Figure S7(a)).

To analyze the optimization progress, we have performed principal component analysis (PCA) to reduce dimensionality such that the descriptor values for a given alloy composition can be visualized in the three dimensions. We first performed PCA on the original Couzinie *et al.* dataset[34] that was used to train and validate the model (Figure S7). Using the same PCA transformation, Figure S8 shows the values of the principal components for both the 25 °C and the 1,000 °C optimal alloy along the optimization path. Considering first the 25 °C optimal alloy in Figure S8a, the optimization proceeds in the direction of increasing yield strength, ultimately optimizing primarily through an increase in the value of the second and third principal component. Evident from the PCA loading plot in Figure S7 is that the second principal component (PC2) is primarily composed of positive loading of $x_{Mo}$, whereas the third principal component (PC3) mainly of positive loadings of $x_{Mo}$, $\delta$, and $\delta G_{Ta}$. Optimization occurring through an increase in PC2 and PC3 is, therefore, consistent with the observation made in the Manuscript for Figure S7 that an increase in $x_{Mo}$ is frequently correlated to an improvement of yield strength. For the optimization trajectory performed at 1,000 °C in Figure S8b, the initial and final point are noticeably closer together in principal component space, indicating that the base alloy and optimized alloy are relatively similar materials. Nevertheless, the optimal alloy has a noticeably smaller value of PC3, which corresponds in an observed decrease of δ. We also want to emphasize that the protocol presented here does not sample the composition space randomly to find the composition with a maximized yield strength, but instead maximizes the yield strength through an intelligent navigation of the composition space informed by previous iterations.

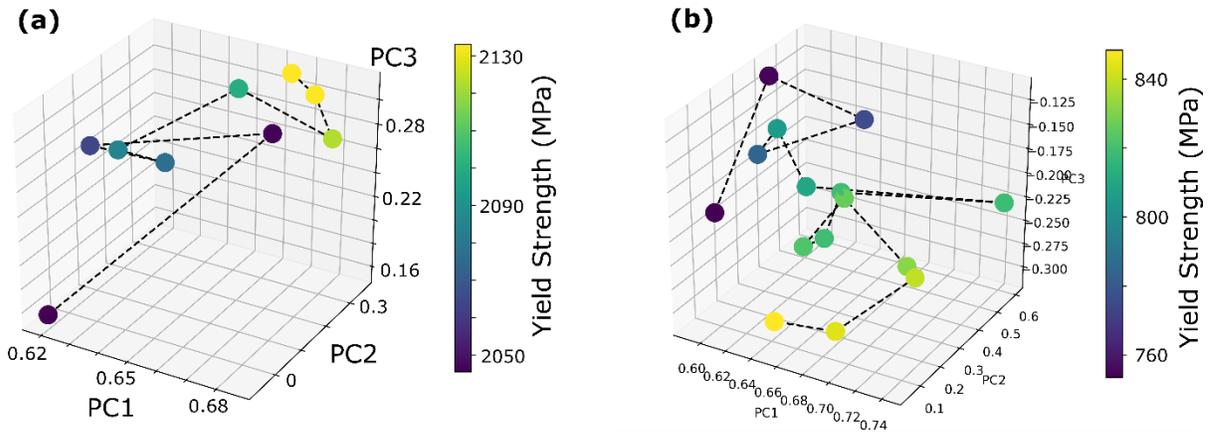

*Figure S8: Optimization trajectories in principal component space for the (a) 25 °C optimization and the (b) 1,000 °C optimization of the AlMo$_{0.5}$NbTa$_{0.5}$TiZr base alloy.*

**Example Calculation of Yield Strength Using Varvenne and Curtin Model**

We have utilized the temperature-dependent yield strength model formulated by Varvenne and Curtin[45] to augment the experimentally available HEA dataset at temperatures between 25 °C and 600 °C. The accuracy of ML approaches is always fundamentally limited by the availability of adequate data, and HEA mechanical properties at temperatures between 25 °C and 600 °C are typically not reported. Using the experimental yield strengths at 25 °C and 600 °C and the temperature-dependent yield strength model proposed by Varvenne and Curtin[45], the yield strength for 200 °C and 400 °C were computed. An example calculation is provided below for the AlMo$_{0.5}$NbTa$_{0.5}$TiZr HEA, which has a yield strength of 2000 MPa at 25 °C and 1870 MPa at 600 °C.

The thermally-activated finite-temperature yield strength is given by Varvenne and Curtin as,

$$\tau_y(T, \dot{\varepsilon}) = \tau_{y0}\exp\left(-\frac{1}{0.51}\frac{kT}{\Delta E_{b0}}\ln\frac{\dot{\varepsilon}_0}{\dot{\varepsilon}}\right)$$

, where $\tau_{y0}$ is the zero-temperature flow stress, $k$ is the Boltzmann constant, $T$ is the temperature, $\Delta E_{b0}$ is the dislocation energy barrier, $\dot{\varepsilon}$ is the strain rate (a value that is always reported experimentally when generating the stress-strain curve), and $\dot{\varepsilon}_0$ is the reference strain rate (taken as $10^4$ s$^{-1}$). Let $T_1 = 25$ °C = 298.15 K and $T_2 = 600$ °C = 873.15 K, where $\tau_y(T_1, \dot{\varepsilon})$ and $\tau_y(T_2, \dot{\varepsilon})$ are known experimentally. Then, $\tau_{y0}$ cancels when calculating the ratio of the two known yield strengths:

$$\frac{\tau_y(T_1, \dot{\varepsilon})}{\tau_y(T_2, \dot{\varepsilon})} = \frac{\exp\left(-\frac{1}{0.51}\frac{kT_1}{\Delta E_{b0}}\ln\frac{\dot{\varepsilon}_0}{\dot{\varepsilon}}\right)}{\exp\left(-\frac{1}{0.51}\frac{kT_2}{\Delta E_{b0}}\ln\frac{\dot{\varepsilon}_0}{\dot{\varepsilon}}\right)}$$

Thus, the only unknown is the zero-temperature energy barrier, $\Delta E_{b0}$. Taking the natural logarithm of both sides and solving for $\Delta E_{b0}$ yields:

$$\Delta E_{b0} = \frac{1}{\ln(\frac{\tau_y(T_1, \dot{\varepsilon})}{\tau_y(T_2, \dot{\varepsilon})})}\left(-\frac{1}{0.51}k(T_1 - T_2)\ln\frac{\dot{\varepsilon}_0}{\dot{\varepsilon}}\right)$$

The yield strength measurements for the AlMo$_{0.5}$NbTa$_{0.5}$TiZr HEA were performed at a strain rate of $10^{-3}$ s$^{-1}$. Therefore, $\Delta E_{b0}$ is calculated from the experimental data as:

$$\Delta E_{b0} = \frac{1}{\ln(\frac{2000 \text{ MPa}}{1870 \text{ MPa}})}\left(-\frac{1}{0.51}(8.6713 \times 10^{-5} \text{ eV K}^{-1})(298.15 - 873.15 \text{ K})\ln\frac{10^4 \text{ s}^{-1}}{10^{-3} \text{ s}^{-1}}\right)$$

$$\Delta E_{b0} = 21.74 \text{ eV} = 2085 \text{ kJ mol}^{-1}$$

The value of $\Delta E_{b0}$ calculated from the experimental yield strength values can then be used to calculate the yield strength at a new temperature, $T$. Calculating for $T = 400\,°C$:

$$\tau_y(T, \dot{\varepsilon}) = \tau_y(T_1, \dot{\varepsilon}) \left( -\frac{1}{0.51} \frac{k(T - T_1)}{\Delta E_{b0}} \ln \frac{\dot{\varepsilon}_0}{\dot{\varepsilon}} \right)$$

$$\tau_y(T, \dot{\varepsilon}) = 2000\text{ MPa} \left( -\frac{1}{0.51} \frac{(8.6713 \times 10^{-5}\text{ eV K}^{-1})(673.15\text{ K} - 873.15\text{ K})}{21.74\text{ eV}} \ln \frac{10^4\text{ s}^{-1}}{10^{-3}\text{ s}^{-1}} \right)$$

$$\tau_y(T, \dot{\varepsilon}) = 1959.5\text{ MPa}$$

**Repetition of Feature Selection Process**

Sequential feature selection (SFS)[2,3] was performed 25 times to assess the variability of which features are included in the model. The result of the SFS replication is provided in Figure S10. The test temperature is selected as one of the six features in every model. Yield strength has a well-known inverse relationship with the temperature as materials generally weaken with increasing temperature. The relationship of yield strength with the other composition-based descriptors are generally less understood. From the SFS replication, the most frequently selected descriptors, in addition to test temperature, were: dG Mo (molybdenum modulus distortion), Mo (element fraction of molybdenum), dr Al (lattice distortion of aluminum), Yang VEC (valence electron concentration), Yang HPsigma_min05 (half the average base strength, plus half the minimum elemental base strength), Yang omega (defined in Equation 4 of the Manuscript), Yang delta (defined in Equation 1 of the Manuscript), dG Ta (modulus distortion of tantalum), Al (aluminum element fraction), Anneal (0/1 hot-encoded variable indicated whether the synthesized alloy was annealed following casting), and Yang delXi (the standard deviation of the electronegativity).

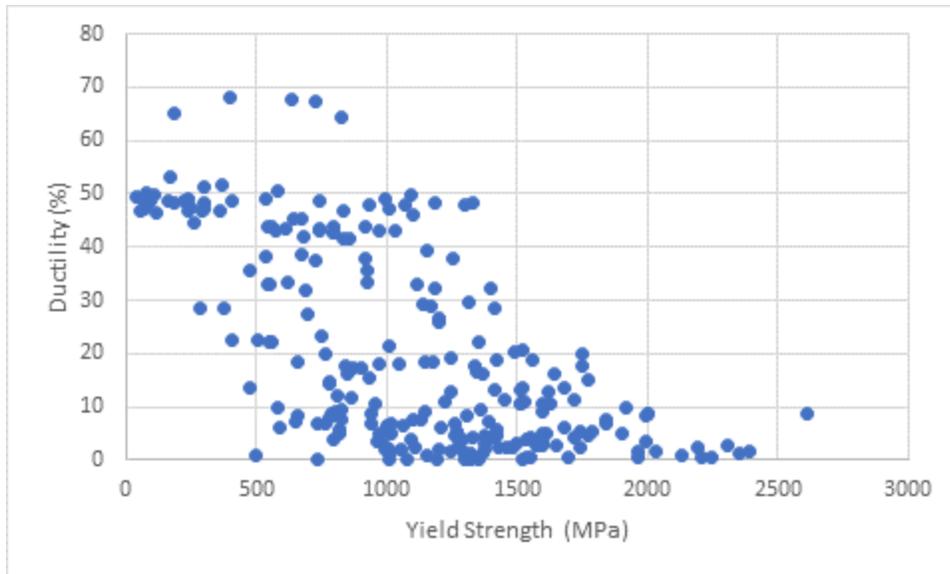

*Figure S9. Experimental values of yield strength and ductility at different temperatures from the Couzinie et al. dataset[1].*

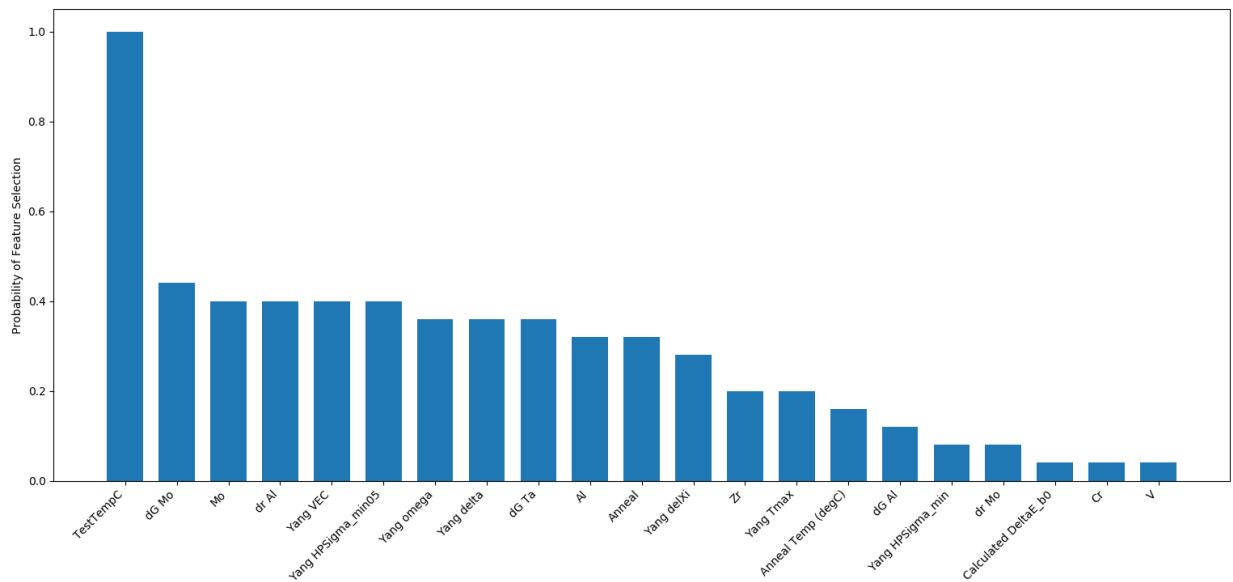

*Figure S10. Histogram of feature selection probability for a six-feature random forest model.*

**References**

1. Couzinié, J. P., Senkov, O. N., Miracle, D. B. & Dirras, G. Comprehensive data compilation on the mechanical properties of refractory high-entropy alloys. *Data Br.* **21**, 1622–1641 (2018).

2. Pudil, P., Novoviĉová, J. & Kittler, J. Floating search methods in feature selection. *Pattern Recognit. Lett.* **15**, 1119–1125 (1994).

3. Ferri, F. J., Pudil, P., Hatef, M. & Kittler, J. Comparative study of techniques for large-scale feature selection. *Pattern Recognit. Pract. IV* 403–413.